\documentclass{article}

\usepackage{mlspconf}

\usepackage{cite}
\usepackage{amsmath,amssymb,amsfonts,amsthm}
\usepackage{algorithmic}
\usepackage{graphicx}
\usepackage{textcomp}
\usepackage{tabularx}
\usepackage{multirow}
\usepackage{multicol}
\usepackage{comment}
\usepackage[dvipsnames]{xcolor}
\usepackage{float}
\usepackage{tikz}
\usetikzlibrary{calc}
\usepackage{stackengine}
\usepackage{todonotes}
\tikzset
  {main node/.style=
    {rectangle,fill=white!5,draw,minimum size=0.5cm,inner sep=2pt}
  } 

\theoremstyle{remark}
\newtheorem{remark}{Remark}
\theoremstyle{definition}

\def\BibTeX{{\rm B\kern-.05em{\sc i\kern-.025em b}\kern-.08em
    T\kern-.1667em\lower.7ex\hbox{E}\kern-.125emX}}
    
\usepackage{pgfplots}
\pgfplotsset{compat=1.3}
\pgfplotsset{
    discard if/.style 2 args={
        x filter/.code={
            \edef\tempa{\thisrow{#1}}
            \edef\tempb{#2}
            \ifx\tempa\tempb
                
            \fi
        }
    },
    discard if not/.style 2 args={
        x filter/.code={
            \edef\tempa{\thisrow{#1}}
            \edef\tempb{#2}
            \ifx\tempa\tempb
            \else
                
            \fi
        }
    }
}

\copyrightnotice{979-8-3503-2411-2/23/\$31.00 {\copyright}2023 IEEE}

\toappear{2023 IEEE International Workshop on Machine Learning for Signal Processing, Sept.\ 17--20, 2023, Rome, Italy}

\title{Coded Distributed Image Classification}

\name{Jiepeng Tang$^{\dagger}$, \qquad Navneet Agrawal$^{\ddagger}$, \qquad Slawomir Stanczak$^{\ddagger\star}$, \qquad Jingge Zhu$^{\dagger}$%
    \thanks{JT and JZ acknowledge the support by the Australian Research Council under Project DE210101497 and the UoM-BUA Partnership scheme. NA is funded by the German Research Foundation (DFG) within their priority program SPP1914 ``Cyber-Physical Networking'', and SS acknowledges the support of joint project 6G-RIC with project identification numbers 16KISK020K and 16KISK030. (email: navneet.agrawal@tu-berlin.de)}}

\address{$^{\dagger}$ University of Melbourne, Australia, \quad
        $^{\ddagger}$ Technische Universität Berlin, Germany, \\
        $^{\star}$ Fraunhofer Heinrich Hertz Institute, Germany}

\newcommand{\Field}[1]{\mathbb{#1}}

\newcommand{\Set}[1]{\mathcal{#1}}
\newcommand{\Operator}[1]{\mathsf{#1}}

\newcommand{\Sseq}[2]{\{#1, \dots ,#2\}}

\newcommand{\Matrix}[1]{\mathbf{#1}}
\newcommand{\Vector}[1]{\pmb{#1}}

\newcommand{\Vect}{\text{Vec}}

\newcommand{\Real}{\Field{R}}
\newcommand{\Realp}{\Real_+}

\newcommand{\Natural}{\Field{N}}

\newcommand{\sS}{\Set{S}}
\newcommand{\sF}{\Set{F}}
\newcommand{\sX}{\Set{X}}

\newcommand{\sD}{\Set{D}}

\newcommand{\opf}{\Operator{f}}

\newcommand{\opE}{\Operator{E}}

\newcommand{\opH}{\Operator{H}}

\newcommand{\opv}{\Operator{D}}

\newcommand{\opD}{\Gamma}

\newcommand{\opL}{\Lambda}

\newcommand{\mX}{\Matrix{X}}

\newcommand{\mV}{\Matrix{V}}

\newcommand{\tX}{\tilde{\mX}}

\newcommand{\ve}{\Vector{e}}

\newcommand{\vy}{\Vector{y}}
\newcommand{\vf}{\Vector{f}}

\newcommand{\hf}{\hat{\vf}}
\newcommand{\ty}{\tilde{\vy}}
\newcommand{\ay}{\acute{\vy}}

\newcommand{\psc}{AICC}

\newcommand\blankfootnote[1]{%
  \let\svthefootnote\thefootnote%
  \let\thefootnote\relax\footnotetext{#1}%
  \let\thefootnote\svthefootnote%
}

\begin{document}
\ninept
\maketitle



\begin{abstract}
In this paper, we present a coded computation (CC) scheme for distributed computation of the inference phase of machine learning (ML) tasks, specifically, the task of image classification.
Building upon Agrawal et al.~2022, the proposed scheme combines the strengths of deep learning and Lagrange interpolation technique to mitigate the effect of straggling workers, and recovers approximate results with reasonable accuracy using outputs from any $R$ out of $N$ workers, where $R\leq N$.
Our proposed scheme guarantees a minimum recovery threshold $R$ for non-polynomial problems, which can be adjusted as a tunable parameter in the system. 
Moreover, unlike existing schemes, our scheme maintains flexibility with respect to worker availability and system design.
We propose two system designs for our CC scheme that allows flexibility in distributing the computational load between the master and the workers based on the accessibility of input data.
Our experimental results demonstrate the superiority of our scheme compared to the state-of-the-art CC schemes for image classification tasks, and pave the path for designing new schemes for distributed computation of any general ML classification tasks.
\end{abstract}


\section{Introduction} \label{sec:intro}



    Cloud platforms provide distributed computation services over many computing nodes for large-scale tasks. 
    However, distributed computing systems are susceptible to straggler effects caused by slow or failing nodes, which can result in high latency in computation tasks. 
    A na\"ive approach to address this issue is replication, which involves executing task copies on several computing nodes (workers) and returning the fastest-responding clones' results. 
    However, this method incurs significant resource overhead. 
    In this direction, a class of techniques known as coded computation (CC) has emerged as a prominent alternative.
    In CC schemes, redundancy is introduced into the input data based on the coding theory, and the encoded data is distributed to the workers for computation. 
    This enables a master node to recover full computation results by decoding the results from only a few worker outputs, without waiting for all workers to return their results.
    Thus, such a scheme requires fewer resources to deal with the straggler effects compared to replication-based approaches \cite{Songze2020}.

    Many existing CC schemes excel at recovering exact results on distributed tasks, but they are generally tailor-made for specific classes of functions, such as matrix multiplication\cite{yu2017polynomial,dutta2019optimal}.
    To provide a trade-off between accuracy and recover-threshold, some researchers have proposed CC schemes for approximate recovery of results \cite{gupta2018oversketch,jahani2021codedsketch}.
    A notable example of CC scheme that covers a general class of functions, namely, polynomial functions, is the \textit{Lagrange coded computation} (LCC)\cite{yu2019lagrange}.
    The LCC uses Lagrange's interpolation technique to guarantee exact recovery of results using outputs from a fixed minimum number of workers (known as the recovery threshold of scheme) \cite{Songze2020}.
    However, all above-mentioned CC schemes are severely restricted in their application to several popular machine learning (ML) tasks, which typically incorporate deep neural networks (DNNs) with complex non-linear structures.
    
    Several papers have proposed CC schemes for the training stage of deep neural networks (DNNs),\footnote{A supervised machine-learning algorithm consists of a \emph{training} phase, where model parameters are trained using a dataset, and an \emph{inference} or testing phase, where the pre-trained model is implemented over the new inputs, providing approximate results.}~addressing the problem of stragglers as well as malicious attacks (seminal works include \cite{chen2018draco, amiri2019computation}). 
    For the inference stage of ML tasks, \cite{kosaian2018learning,kosaian2020learning} propose a CC scheme that ``learns'' an appropriate erasure code via a supervised deep learning technique.
    One approach jointly trains an auto-encoder for encoding and decoding operations, while the other trains a deep neural network (DNN) to perform the computation on the encoded data.
    However, they are limited to a single straggler and their performance degrades significantly with an increasing number of inputs.
    
    In \cite{agrawal2022learning}, the authors propose a CC scheme called \psc, that addresses the issues of exiting CC schemes by combining the powers of deep learning and Lagrange's interpolation technique, to provide approximate results in distributed computation tasks.
    The \psc~is (1) applicable to general functions, including non-polynomials; (2) guarantee a recover-threshold that does not rely on the number of inputs; and (3) allows for a judicious trade-off between the accuracy and the computation load.
    However, as a proof-of-concept in \cite{agrawal2022learning}, the \psc~is applied only to certain non-polynomial matrix functions.

    In this paper, we present a CC scheme based on \psc~for the distributed computation of tasks pertaining to the inference phase of ML algorithms.
    Specifically, we apply the proposed CC scheme to an image classification task on the Fashion-MNIST dataset \cite{xiao2017fashion}.
    Unlike existing CC schemes for image classification \cite{kosaian2018learning,kosaian2020learning}, our scheme allows for distribution of the tasks to any number of workers available, while still inheriting the recovery-through-interpolation property of LCC to guarantee the recovery of results from a fixed (but unspecified) subset of worker outputs.
    Furthermore, for applications with privacy concerns of the input data, we present a novel variant of AICC's computation operation that, unlike the scheme proposed in \cite{agrawal2022learning}, only requires the encoded data, while showing only a small degradation in performance.
    Another promising characteristic of the proposed scheme is that most of its design parameters are tunable, making it amenable to the system's capabilities, and trade-off between desired accuracy and available computation resources.
    The experimental results of this study verify that our scheme, using the same DNN architectures as used in \cite{kosaian2020learning}, outperforms the accuracy achieved in \cite{kosaian2020learning} on the Fashion-MNIST dataset.

\section{Preliminaries}

    Consider a function $f$ that maps any real-valued square matrix from $\sD\subset\Real^{M\times M}$ to an element of some finite dimensional Euclidean space $\sS$.
    Note that the restriction of domain of $f$ to square matrices is not a limitation of the proposed scheme, but a choice that reflects the application requirements, i.e.~the images used in the classification task are represented as elements in $\sD$.
    Define a set of $K$ input matrices (called \emph{dataset}) $\sX := \{\mX_{1}, \dots, \mX_{K}\}$, and the set of outputs of function $f$ on the dataset $\sX$ as $\sF := \{f(\mX_1), \dots, f(\mX_K)\}$.
    The objective is to obtain $\sF$, given $\sX$, with a minimum delay.
    The system consists of $N \geq K$ workers and a master node, which is responsible for collecting all the results in $\sF$.
    To reduce the delay in obtaining the results, the computation task is distributed among the workers.
    A na\"ive approach is to distribute the $K$ computations in $\sF$ to any subset $K$ of $N$ workers, and wait for them to return their results to the master.
    Typically, such an approach is prone to unexpected delays due to the so-called \emph{straggling effects} such as worker-node failure or slowdown.
    Moreover, some of the available worker resources are not used in this approach.
    The coded computation (CC) refers to a class of distributed computation schemes which use the available resources more efficiently, and can mitigate the straggling effects by using coding techniques to recover all desired results from a subset of worker outputs.
    
    In general, the CC schemes operate by encoding the dataset $\sX$ to generate $N$ encoded data $\tX_n, n=1,\ldots, N$. 
    The worker $n$ performs some computation on the encoded data $\tX_n$ and returns the result to the master.
    Then, the master recovers the desired results $\sF$ upon receiving the computation results from any subset $R \leq N$ of workers. 
    The smallest value of $R$ for which the recovery of results can be guaranteed is called the \textit{recovery threshold} of the scheme.
    In this way, the CC schemes can mitigate the impact of straggling.

    The LCC~\cite{yu2019lagrange} scheme is a CC scheme specifically designed only for polynomial functions $f$, say, of degree $d$.
    The functioning principle of the LCC can be summarized as follows (see~\cite{yu2019lagrange} for details):
    The encoding operation amounts to the evaluation of a Lagrange polynomial at some distinct nonnegative scalars $\alpha_n$ for $n=1,\dots,N$.
    The Lagrange polynomial is of degree $K-1$, and its coefficients are linear functions of the input data $\sX$, making it computationally inexpensive to implement.
    The worker $n$ then computes the polynomial $f$ on the encoded data $\tX_n$.
    The computation result $f(\tX_n)$ can be viewed as a composition of the Lagrange polynomial and the polynomial function $\opf$ at a point $\alpha_n$, and hence, it is a polynomial itself, of total degree $(K-1)d$.
    Hence, by using the Lagrange interpolation technique, this composite polynomial can be recovered (exactly) using $R = (K-1)d +1$ results from the workers.
    The desired function values $\sF$ can then be obtained by evaluating the polynomial at suitable points known to the master.
    Although LCC is optimal in terms of the recovery threshold, it suffers from two crucial limitations: 
    (i) the function $f$ is restricted to be a polynomial of the matrix-valued inputs; and
    (ii) the recovery threshold $R$ of LCC grows proportionally with the number of inputs $K$ and degree $d$ of the polynomial $f$.

    In \cite{agrawal2022learning}, a learning-based approximate CC scheme is proposed, called \psc, to tackle the limitations of the LCC.
    The \psc~scheme can be applied to a large class of functions, including non-polynomial functions, and provide approximate results which can be tuned to a desired accuracy based on the application requirements and system capabilities.
    The \psc~scheme is shown in \cite{agrawal2022learning} to provide reasonably accurate results for certain matrix-valued functions, namely, computation of eigenvalues, dominant eigenvector, determinant, and exponential of a matrix, that are of interest in wireless communications.

    The main objective of this paper is to extend the \psc~scheme to functions that amount to evaluation of the inference stage of an ML algorithm.
    Specifically, we propose a CC scheme, based on \psc, for the problem of \emph{image classification} over a given dataset.
    The ML algorithm for image classification task is structured so that most of its computations in the inference stage can be distributed to the workers.
    By carefully designing the encoding and computation operations such that their composition has a polynomial structure, we ensure that the desired results can be decoded from any subset of $R$ worker results, as in the \psc~scheme.
    It is worth noting that the image classification task considered in this paper requires different design approach compared to the approximate function computation tasks tackled in \cite{agrawal2022learning}.
    For example, the output of the function in image classification problem is the predictive probability of a certain class, while the target is one of the classes.
    Hence, the learning framework must be modified to support the problem and the available data.
    Moreover, the learning capabilities of DNNs involved in \psc~can be enhanced by embedding the knowledge about the problem into design of DNNs.

    As \psc~is an essential part of this study, we dedicate Section \ref{sec:AICC} to its introduction.
    In Section \ref{sec:IMG}, we formulate the image classification problem, provide details of the design choices for the DNNs involved in \psc~encoder and computation operations, and describe the training procedure that achieves desired accuracy of results.
    The details of simulation, and the results comparing the proposed method with existing approaches, are provided in Section \ref{sec:sim}.

\section{AI-aided Coded Computation (\psc)} \label{sec:AICC}

    In this section, we give a brief overview of the \psc~scheme as proposed in \cite{agrawal2022learning}.
    Similar to the LCC scheme, the basic operations involved in the \psc~scheme are \emph{encoding}, \emph{computation}, and \emph{decoding} operations.
    The \psc~enjoys the recover-through-interpolation property of the LCC, with guarantees of a recovery threshold that, in contrast to LCC, depends solely on tunable design parameters.
    At its core, the \psc~scheme involves DNNs into its operations, whose learnable parameters are trained a priori using a training dataset.
    Indeed, the design and architecture of the DNNs in \psc~influences its overall performance, and hence, they need to be carefully designed for a given problem.
    Most importantly, the recover-through-interpolation property of \psc~rely on the polynomial structure of composition of \psc~operations.
    This design aspect is highlighted in this section.

    \subsection{Encoding operation $\opE$} \label{sec:enc}
    In CC schemes, encoding injects redundancy into the data before it is distributed to workers, aiming to obtain desired results with simple decoding. 
    For example, LCC employs a linear function for encoding, resulting in a simple linear decoder \cite{yu2019lagrange}. 
    Given the dataset $\sX \in (\sD)^K$, the encoding operation in \psc~is is a degree $G$ polynomial $\opE : \Realp \to \Real^{M \times M}$ given by $\opE(\alpha) := \sum_{g=0}^G C_g \alpha^g$, where the coefficients $C_g \in \Real^{M\times M}$, for $g=0,\dots,G$, are generated using the function $\opD_g:(\sD)^K \to \Real^{M \times M}$, that involves pre-trained DNNs taking $\sX$ as input. 
    Each worker $n=1,\dots,N$ obtains the matrix $\tX_n = \opE(\alpha_n)$, where $\alpha_n>0$ is a distinct for each worker $n$, and it is known to the master for decoding.
    We remark that the coefficients $(C_g)$, or equivalently the functions $(\opD_g)$, are identical for all workers.
    It is also worth noting that DNNs involved in functions $(\opD_g)$ can be arbitrary, as they do not affect the polynomial structure of the encoder $\opE$.
    Specific DNN designs for $(\opD_g)$ for the image classification problem will be discussed in Section \ref{sec:IMG}.

    \subsection{Computation operation $\opH$} \label{sec:comp}
    The computation is performed by each worker $n$ on the encoded data $\tX_n := \opE(\alpha_n)$, and the results are sent to the master as soon as they become available.
    The computation operation $\opH : \Real^{M\times M} \to \Real^V$ is given by $\opH(\mX) := \sum_{p=0}^P \mV_p\ \Vect(\mX^p)$, where $\Vect(\mX)$ vectorizes the matrix $\mX$ by stacking its columns, and the coefficients $\mV_p := \opL_p(\sX)$, for $p=0,\dots,P$, are generated using the function $\opL_p:(\sD)^K \to \Real^{V\times M^2}$.
    The output dimension $V$ and the degree $P\in\Natural$ are design parameters.
    
    \subsection{Decoding operation} \label{sec:dec}
    The decoding operation essentially involves the following two steps:
    First, the coefficients of the composite polynomial $\opv := \opH \circ \opE : \Realp \to \Real^V$ of degree $GP$ are obtained using the Lagrange interpolation technique via any $R=GP+1$ input-output pairs from the set of worker outputs $\{(\alpha_1, \opv(\alpha_1)), \dots, (\alpha_N, \opv(\alpha_N))\}$, where $R \leq N$ is one more than the degree of the polynomial $\opv$.
    Then, the polynomial $\opv$ is evaluated at $K$ distinct scalars $\beta_1, \dots, \beta_K$ to obtain the desired results.
    Each scalar $\beta_k$, $k=1,\dots,K$, is chosen prior to the training procedure (as described in the following section), such that the output $\opv(\beta_k)$ either corresponds to the approximate result $\hat{f}(\mX_k) \approx f(\mX_k)$ itself, or it can be easily transformed into one.
    An advantage of using the Lagrange's interpolation technique for the decoding operation is that both steps, the polynomial interpolation and the evaluation of results, involve only linear matrix operations, which are computationally inexpensive.

    \begin{remark}[Placement of operations] \label{remark1}
        In systems where the dataset $\sX$ is available to all worker (e.g.~shared memory), the workers can essentially implement both the encoding and the computation operations, and send the results, along with corresponding $\alpha_n$, to the master.
        In this case, the master only needs to perform the computationally inexpensive decoding operation.
        However, when the dataset $\sX$ is not available to the workers, the master must also perform the encoding operation.
        In addition, the dataset $\sX$ or coefficients $(\mV_p)$ must be shared among the workers.
        Instead, to save communication and computation resources, we present a novel design of the computation operation $\opH$ that only relies on the encoded data $\tX_n$ (see Section \ref{sec:opHb}).
        In Section \ref{sec:sim}, we present CC schemes for both cases described above, and compare the performance with different design parameters.
    \end{remark}

    \subsection{Training procedure} \label{sec:training}

    The parameters of DNNs involved in encoding and computation operations in the \psc~scheme are trained such that, for all $k\in\Sseq{1}{K}$, the cost of approximating the true value $f(\mX_k)$ with the approximate output of the CC scheme $\hf_k := \opv(\beta_k)$ over the training dataset is minimized.
    The distinct real-valued scalars $\beta_k$, for $k=1,\dots,K$, are chosen at the beginning, and they remain fixed during the entire training and test phases.
    In \cite{agrawal2022learning}, the mean square loss function is used for the training.
    The training consists of forward and backward passes.
    In each forward pass, for each $k=1,\dots,K$, for every data pair $(\sX, \beta_k)$, the output $\hf_k := \opv(\beta_k) = \opH(\opE(\beta_k))$ is obtained.
    Then, in the backward pass, the parameters are updated based on the loss function between the output $\hf_k$ and the target value $f(\mX_k)$.


\section{Coded Distributed Image Classification} \label{sec:IMG}
    
    The DNN based ML approaches to the image classification problems are known to perform very well \cite{kadam2020cnn}.
    However, for most problems, the DNNs that provide sufficiently accurate results consist of a large number of parameters, and hence, their implementation requires significant computational resources.
    Moreover, in many applications, the image classification problem is required to be solved for numerous images at once, and with minimum latency.
    Therefore, distributed computation is often desirable in such applications to reduce latency.

    In this section, we propose a CC scheme based on \psc~that allows distributed computation of the image classification tasks.
    We describe the encoding, computation, and decoding operations in Sections \ref{sec:img_enc}, \ref{sec:img_comp}, and \ref{sec:img_dec}, respectively.
    In addition to ensuring the polynomial structure of the composite function $\opv:=\opH\circ\opE$ in our design, we also adopt DNN architectures that are known to enhance learning over image data.
    We compare our scheme with the state-of-the-art CC scheme proposed in \cite{kosaian2018learning} for such tasks, and present results from conventional (centralized) DNN-based approaches as a benchmark.
    We remark that, in addition to being superior in performance compared to \cite{kosaian2018learning}, our scheme is inherently more flexible in terms of system requirements (i.e.~accuracy and complexity of the algorithm, as well as recovery threshold, are design parameters), and guarantee a fixed recovery threshold for any number of workers.

    \subsection{Encoding operation design} \label{sec:img_enc}

        The encoding operation follows the same structure as described in Section \ref{sec:enc}, i.e.~$\opE$ is a polynomial of degree $G$, with pre-trained coefficients $C_g := \opD_g(\sX)$, for all $g=0, \dots, G$.
        The two architectures employed for the DNNs in function $\opD_g$, for all $g=0,\dots,G$, are given in Table \ref{table:enc} (activation functions are omitted from the Table).
        One involves the multilayer perceptrons (MLPs), which makes use of fully-connected (FC) NN layers, and another is based on the convolution layers (CLs), which are known to perform well on image data.
        Both architectures use the ReLU activation functions in all but the final layer, and CLs use dilated convolution, as described in \cite{yu2015multi}.
    
        \subsubsection{MLP for encoding polynomial $\opE$}
        For all $g=0,\dots,G$, the coefficient $C_g$ in the encoding operation $\opE$ is output of the function $\opD_g$.
        The function $\opD_g$ is an MLP with the same architecture, but with different learned parameters.
        For MLP in $\opD_g$, the input is the $KM^2$ dimensional vector obtained by flattening each image in the dataset $\sX$ and concatenating them together.
        The output of $\opD_g$ (dimension $M^2$) is transformed into a $M\times M$ matrix to obtain the $g$th coefficient $C_g$ of the encoding polynomial.
        Note that the learnable parameters in the MLP grow proportional to the size of the input $KM^2$, which could lead to overfitting in ML tasks \cite{kosaian2020learning}.
            
            


  \begin{table}[!htbp]
  \centering
  \caption{DNN architectures used in the proposed CC scheme.}
  \label{table:enc}
    \begin{tabular}{c|c|c}
    \hline
    MLP                        & CLs                      & Base-MLP \\ \hline
    $KM^2\times KM^2$          & kern:3$\times$3, dilation 1             & $M^2 \times L_1$ \\
    $KM^2\times M^2$           & kern:3$\times$3, dilation 1             & $L_1 \times L_2$ \\
                               & kern:3$\times$3, dilation 2             & $L_2 \times V$ \\ 
                               & kern:3$\times$3, dilation 4             & \\
                               & kern:3$\times$3, dilation 8             & \\
                               & kern:3$\times$3, dilation 1             & \\
                               & kern:3$\times$3, dilation 1             & \\ \hline
    \end{tabular}
\end{table}
    
        \subsubsection{Convolution layers (CLs) for encoding polynomial $\opE$} \label{sec:en_CNN}

        The CL has the advantage of processing the image in its original 2D structure, and hence, it can learn complex patterns in the image that are usually lost when the image is flattened.
        The CLs, representing the function $\opD_g$ for $g=0, \dots, G$, takes $K$ images in the dataset $\sX$ as $K$ input channels in the first CL.
        In every CL, a kernel of dimension $3\times 3$ is used with dilation (see \cite{yu2015multi}), and between each pair of CLs, there is a ReLU activation function (omitted in the table \ref{table:enc}).
        The dilation increases the receptive field of the image without increasing parameters, captures features at multiple scales, and reduces spatial resolution loss compared to regular convolutions with larger filters \cite{yu2015multi}.
        Even though the DNN architecture with CLs requires more layers to combine all input pixels, it consists of fewer parameters compared to the MLP architecture, since the parameters are only required for defining the $3\times 3$ kernel in a CL.
        This helps the CL architecture to be computationally inexpensive, and also avoid overfitting.
        Hence, employing CL architecture provides a better performance than MLP architecture on the image classification task, which is also confirmed via simulation in Section \ref{sec:sim}.

    \subsection{Computation operation design} \label{sec:img_comp}
    For the computation operation $\opH$, as described in \ref{sec:comp}, we set the design parameter $V$ as the number of classes in the image classification task.
    The worker computation $\opH$ must be carefully designed to ensure that each component of its output $\opH(\tX)=\opH(\opE(\alpha))\in\Real^V$ is a polynomial in $\alpha$.
    In light of remark \ref{remark1}, we propose two designs for the computation function $\opH$ in the following: 
    the first design $\opH_S$, as described in \ref{sec:comp}, requires the dataset $\sX$ as input, and the second, $\opH_B$ is designed to implemented without $\sX$.
    
    

    \subsubsection{Computation function $\opH_S$}
\begin{figure}[htbp]
    \centering
    \includegraphics[width=0.9\columnwidth]{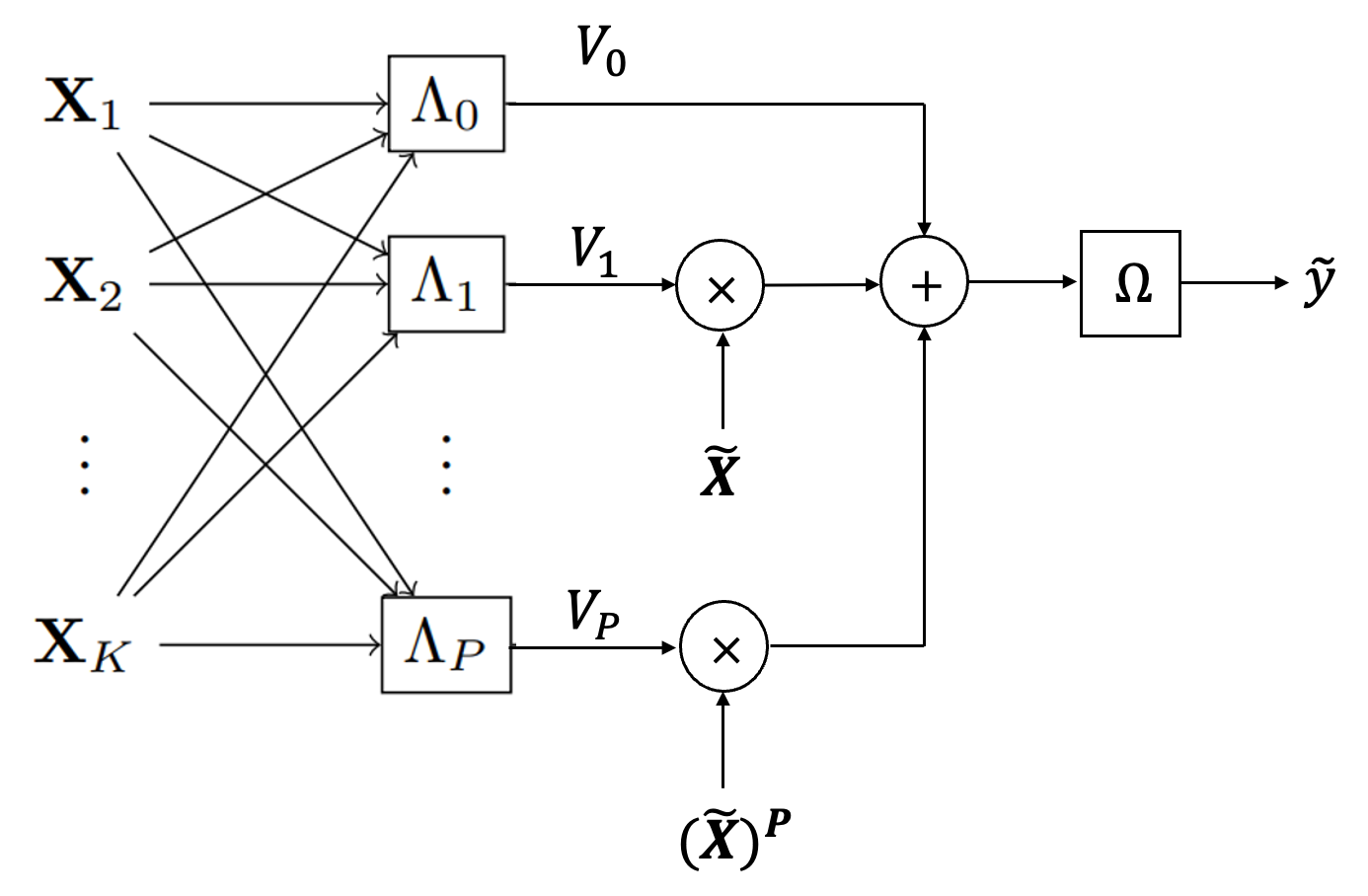}
    \caption{The computation function $\opH_S$ has a polynomial structure whose coefficients are functions of the input dataset $\sX$. On the other hand, the function $\opH_B$ has a linear structure, and it does not depend on the input dataset $\sX$.}
    \label{fig:computationhs}
    \end{figure}

    The computation function $\opH_S: \Real^{M \times M} \to \Real^V$, illustrated in Figure \ref{fig:computationhs}, is defined as follows: 
    \begin{align} \label{eq:comphs}
        \opH_S(\tX) := \Omega\left(\sum_{p=0}^P \mV_p\ (\tX^{\odot p})\right), 
    \end{align}
    where $P\in\Natural$ is a design parameter. 
    The coefficients $\mV_p \in \Real^{M\times M}$, $ p = 0, \dots, P$, are obtained via the functions (involving DNNs) $\opL_p:(\sD)^K\to\Real^{M \times M}$ such that $\mV_p:=\opL_p(\sX)$. 
    The notation $\tX^{\odot p}$ denotes the output of taking $p$ times the Hadamard or element-wise product of the matrix $\tX$ with itself.
    Note that the Hadamard product does not modify the polynomial structure of the output with respect to $\alpha$ in components of the encoded matrix $\tX$, but raises the degree by $p$ times.
    For the function $\opL_p$, we use the same MLP and CL architecture (including the activations) as in function $\opD$ of the encoding operation, as shown in Table \ref{table:enc}.
    The function $\Omega$, defined as $\Omega:\Real^{M\times M} \to \Real^V$, has the Base-MLP architecture (with no activation functions) in Table \ref{table:enc} with parameters $L_1=200$ and $L_2=100$.
    In this design, it is assumed that either the dataset $\sX$ is available to the workers, or the master node (with access to the dataset) computes coefficients $\mV_p$ and broadcast them to the workers (coefficients $(\mV_p)$ are the same for all workers).
    

    \subsubsection{Computation function $\opH_B$} \label{sec:opHb}
        The computation function $\opH_B$ is designed to only depend on the inputs $\tX$, and hence, workers implementing $\opH_B$ do not need access to the dataset $\sX$ or any other information from the master, except the encoded input $\tX$.
        The function $\opH_B:\Real^{M\times M} \to \Real^V$ applies a sequence of linear transformations on the input matrix $\tX$, keeping the polynomial structure of the composite function $\opv := \opH \circ \opE$.
        The linear operator $\opH_B$ is essentially a cascade of matrix multiplications on the input $\tX$, and it can be seen as applying a DNN (MLP or CNN), without any non-linear activation functions, to the input $\tX$.
        In other words, the function $\opH_B$ simply applies some linear matrix operations to the input $\tX$, which does not change the degree of the composite polynomial.
        

        We employ two DNNs architectures for the function $\opH_B$ as shown in TABLE \ref{table:enc}: (1) \emph{MLP}-$\opH_B$: the Base-MLP only and (2) \emph{CNN}-$\opH_B$: CLs followed by the Base-MLP. 
        Note that both architectures {MLP}-$\opH_B$ and {CNN}-$\opH_B$ are without any activation functions.

        \vspace{1em}

\begin{figure}[htbp]
        \centering
        \includegraphics[width=0.9\columnwidth]{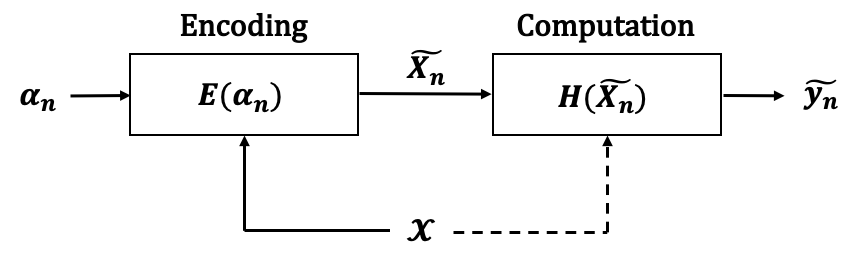}
        \caption{Encoding and Computation process}
        \label{fig:computationhs_enccom}
        \end{figure}

        Figure \ref{fig:computationhs_enccom} illustrates the general data transformation process of our approach during the encoding and computation operations, where the function $\opH$ could be either $\opH_S$ or $\opH_B$.
        The dotted line connecting the input dataset $\sX$ and function $\opH$ represents the two design strategies $\opH_S$ and $\opH_B$, indicating whether the input dataset $\sX$ is used in the computation operation or not.
        As a baseline for comparison, we use two \emph{centralized} DNN based algorithms for the image classification task:
        (1) \emph{MLP-baseline}, and (2) \emph{CNN-baseline}, by respectively adding ReLU activation function between each pair of layers in MLP-$\opH_B$ and CLs of the CNN-$\opH_B$.
        Note that the baseline algorithm is not distributable, and does not apply any encoding or decoding operation.

    \begin{figure}[htbp]
    \centering
    \includegraphics[width=0.9\columnwidth]{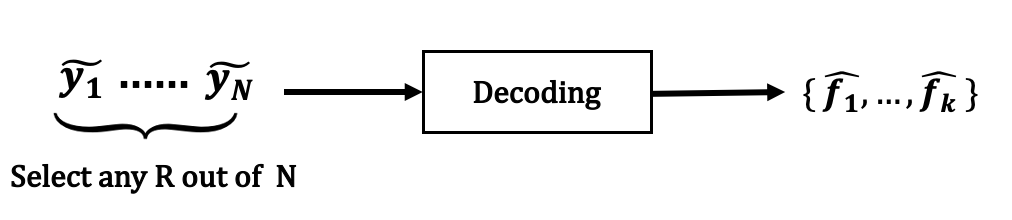}
    \caption{Decoding process}
    \label{fig:computationhs_dec}
    \end{figure}

    \subsection{Decoding operation design} \label{sec:img_dec}
        The decoding operation is essentially the same as described in Section \ref{sec:dec}, except that after recovering, we apply a \emph{Softmax} operation to convert the vector $\ty\in\Real^V$ to a probability mass function for each of the $V$ classes.
        The recovery is ensured by the polynomial structure of the composite polynomial $\opv:=\opH \circ \opE$.
        Using computation operation $\opH_S$ or $\opH_B$ results in the polynomial $\opv$ of degree $GP$ or $G$, respectively.
        Thus, the recovery threshold for the schemes corresponding to $\opH_S$ and $\opH_B$ is $R=GP+1$ and $R=G+1$, respectively.
        After interpolating the polynomial $\opv$, we apply the \textit{Softmax} function to the result $\opv(\beta_k)$ to obtain $\ay_k := \text{Softmax}(\opv(\beta_k))$, and select the index with the highest probability as the prediction label $L_k$ for the input image $\mX_k$.
        



    \subsection{Training procedure} \label{sec:img_training}
        We follow the same training procedure as described in Section \ref{sec:training}, except that the loss function is different in our scheme.
        In the training dataset, we are given multiple data with the inputs $\sX := \{\mX_1, \dots, \mX_K\}$ and corresponding true labels $f(\mX_k)\in\{\ve_1, \dots, \ve_V\}$, for each $v=1,\dots,V$, where $\ve_v\in\Real^V$ is the one-hot vector with all elements zero except one at position $v$.
        At each forward pass of the scheme with input $\beta_k$, where $(\beta_1, \dots, \beta_K)$ are distinct scalars that are fixed, we obtain $\ay_k := \text{Softmax}(\opv(\beta_k))$ via the encoding, computation, and decoding operations, as described above.
        The backward pass updates the DNN parameters using an iterative algorithm based on the \emph{cross entropy loss function} between the true labels $f(\mX_k)$ and the estimated probability mass $\ay_k$.

\section{Simulations} \label{sec:sim}


    

    We implemented our learning-based CC model using PyTorch, which enabled us to experiment with different DNN architectures for the encoding and computation operations. 
    We use the Fashion-MNIST dataset \cite{xiao2017fashion} for the image classification task, which has images of dimension $\Real^{M\times M}$ with $M=28$, and $V=10$ target classes.
    We train our model for 20 epochs in two settings: (1) a batch size of 64 samples with $K=2$ inputs in dataset $\sX$, and (2) a batch size of 32 samples with $K=4$ inputs in dataset $\sX$.
    Each minibatch sample has $K$ images drawn randomly without replacement, and no image was sampled more than once per epoch. 
    We used the Adam optimizer with a learning rate of 0.001 to update the DNN parameters. 
    The scalar values $\{\beta_{1}, \beta_{2}, \dots ,\beta_{K}\}$ are fixed to $\beta_{k}=k/K$ for $k=1, \dots, K$ throughout the experiment. 
    During the testing phase, we fix the scalar values $\{\alpha_{1}, \alpha_{2}, \dots,\alpha_{N}\}$ to $\alpha_{n}= n/(N+1)$ for $n=1,2,\dots,N$.
    We evaluate the performance of our model based on its accuracy in reconstructing the results on a separate test dataset. 
    The \textit{evaluation criteria} is the \textit{accuracy} of predictions on the test dataset, defined as the proportion of correctly predicted labels to the total number of images in the test dataset.

\subsection{Results}
In the following, we denote the proposed CC schemes with the computation functions $\opH_B$ and $\opH_S$ as $\opH_B$ and $\opH_S$, respectively.
We compare the performance of our scheme in Section \ref{sec:res1} for different DNN architectures used in encoding and computation operations, as well as with \cite{kosaian2020learning} and centralized baseline models.
Furthermore, in Section \ref{sec:res2}, we investigate our scheme for different selections of encoder and computation polynomial degrees $G$ and $P$, respectively, and discuss how one can select these free parameters in our scheme to make a judicious trade-off between accuracy and computational complexity.

\subsubsection{Comparison of different DNN architectures} \label{sec:res1}
For fair comparison, in our simulation, we use the architecture of MLPencoder and CNNencoder from \cite{kosaian2020learning} for our encoding and computation operations (except that in $\opH_B$ activations are not implemented).
Table \ref{table:comparison} presents results for a system (the same as in \cite{kosaian2020learning}) with workers $N=3$, input size $K=2$ and recovery threshold $R=2$.
Note that, in both our schemes $\opH_B$ and $\opH_S$, to ensure the recovery threshold $R=2$, we use the degrees $(G=1, P=1)$, respectively, for the encoding and computation operations.
Our CC scheme outperforms the schemes presented in \cite{kosaian2020learning} in terms of accuracy, regardless of whether we use $\opH_B$ or $\opH_S$. 
Moreover, the encoder using CL architecture (see Section \ref{sec:en_CNN}) outperforms the MLP encoder in both models.
Compared to the baselines, the degradation in accuracy of our scheme based on MLP and CNN encoders is around 5\% and 7\%, respectively, which is an acceptable degradation considering the fact that the baseline algorithm cannot be distributed.
Most notably, these results demonstrate that CL architecture performs better than MLP, concurring with the discussion in Section \ref{sec:en_CNN}.

\begin{table}[!htbp]
\centering
\caption{Evaluating accuracy of schemes for $R=2$ and $K=2$.} 
\label{table:comparison}
\begin{tabular}{c|c|c|c|c}
\hline
Enc+Comp    & $\opH_B$  & $\opH_S$  & \cite{kosaian2020learning} model & Baseline               \\ \hline
MLP+MLP & \textbf{83.93} & \textbf{84.34} & 81.96   & \multirow{2}{*}{89.12} \\ \cline{1-4}
CNN+MLP & \textbf{85.81} & \textbf{84.52} & 82.53   &                        \\ \hline
MLP+CNN & 84.11 & 85.37 & --\footnotemark[2]   & \multirow{2}{*}{92.01} \\ \cline{1-4}
CNN+CNN & 85.94 & 86.62 & --\footnotemark[2]   &                       \\ \hline
\end{tabular}
\end{table}
\footnotetext[2]{Some of the results of \cite{kosaian2020learning} cannot be compared directly with our scheme because they correspond to the \emph{ResNet18} model in the computation operation, while our schemes $\opH_B$ and $\opH_S$ uses the CL architecture presented in Table \ref{table:enc} (also see Section \ref{sec:IMG}).
Nonetheless, even though the \emph{ResNet18} model is specifically designed CNN architecture that is well-trained for this problem, our schemes, with a simple training procedure and design, only show $2-4\%$ degradation in performance w.r.t.~\cite{kosaian2020learning}.}




\subsubsection{Performance trade-off in selection of design parameters} \label{sec:res2}
We conduct additional experiments to evaluate the impact of two parameters on the performance of our two models, namely, the degree of composite function $\opv:=\opH \circ \opE$, denoted by $D$ in the following, and input size $K$.
Table \ref{table:input-size} shows results for computation operations $\opH_S$ and $\opH_B$, with $D=GP$ and $D=G$, respectively, where we have implemented CL architecture for both operations (see Section \ref{sec:IMG} for details).
The column `\# of params' denotes the total number of learnable parameters (scaled such that $100\% \equiv 2218197$) in all the DNNs involved in our scheme, and it is one of the indicators of computational complexity of the inference using such model.
Within the same parameter setting for $K$ and $R$, the model $\opH_S$ performs better than the model $\opH_B$, as the former benefit from the nonlinear activations involved in coefficients $\mV_p := \opL_p(\sX)$, for $p=0,\dots,P$, of the computation operation.
Moreover, the accuracy gap between the two models widens as the number of inputs $K$ in the dataset increases, which in turn provides more parameters for learning to $\opH_S$ compared to $\opH_B$.


For a given $K$, increasing the value of $D$ enhances the accuracy, which suggests that one can improve model performance by tuning the parameters $G$ and/or $P$.
Also, increasing the value of $P$ in the computation function $\opH_S$ results in a better performance. 
For instance, for $K=4$ and $D=4$, the accuracy of the model $\opH_S$ is $75.6\%$ with $G=4$ and $P=1$, but it improves to $82.3\%$ with $G=1$ and $P=4$, even though both models have the same number of parameters.
In addition to the improved accuracy, for schemes with the same recovery threshold $R$, a lower value of $G$ (or higher value of $P$) means less computation for the encoding operation, which is usually implemented by the master node.
Hence, for model $\opH_S$, by increasing the degree of computation operation and reducing the degree of encoding operation, we can reduce the overhead of the master as well as improve the accuracy.
However, to improve the accuracy of the model $\opH_B$, we can only increase the value of $G$ in the encoding function $\opE$, which would increase the workload of the master.
Nonetheless, the model $\opH_B$ does not require sharing input data $\sX$ with each worker, which enhances data privacy and minimizes network communication. 
Hence, our scheme provides flexibility of the choice of model, based on the specific system requirements and capabilities.

\begin{table}[!htbp]
\centering
\caption{Varying degrees $G,P$ and number of inputs $K$.}
\label{table:input-size}
\begin{tabular}{c|c|c|c|c|c}
\hline
 $K$ &  $R$ & $\opH_B$/$\opH_S$   & $G$, $P$ & accuracy & \# of params \\ \hline
             &                      & $\opH_B$ & 2, 1   & \textbf{87.3}     & 21 \%    \\ \cline{3-6} 
2          & 3                  & \multirow{2}{*}{$\opH_S$} & 2, 1   & 88.5     & 23 \%    \\ \cline{4-6} 
             &                      &         & 1, 2   & \textbf{88.7}     & 23 \%    \\ \hline
             &                      & $\opH_B$ & 2, 1   & \textbf{71.2}     & 50 \%   \\ \cline{3-6} 
             & 3                  & \multirow{2}{*}{$\opH_S$} & 2, 1   & 71.3     & 75 \%   \\ \cline{4-6} 
             &                      &         & 1, 2   & \textbf{76.8}     & 75 \%   \\ \cline{2-6} 
4          &                      & $\opH_B$ & 4, 1   & \textbf{75.6}     & 80 \%   \\ \cline{3-6} 
             & 5                  &         & 4, 1   & 75.6     & 100 \%   \\ \cline{4-6} 
             &                      & $\opH_S$ & 2, 2   & 78.0     & 86 \%   \\ \cline{4-6} 
             &                      &         & 1, 4   & \textbf{82.3}     & 100 \%  \\ \hline
\end{tabular}
\end{table}

\section{Conclusion} \label{sec:con}

In conclusion, this paper proposes a novel coded computation scheme for distributed computation of inference phase of machine learning algorithms based on \psc~\cite{agrawal2022learning}. 
The proposed scheme allows for the distribution of tasks to any number of workers while ensuring the recovery-through-interpolation property of LCC for an approximate recover of results with desired accuracy.
Moreover, a novel variant of AICC's computation operation is presented to address privacy concerns of input data, with only a small degradation in performance. 
The proposed scheme's tunable design parameters make it amenable to systems with varying capabilities, and allows for a judicious trade-off between accuracy and available computation resources. 
Experimental results for image classification on the Fashion-MNIST dataset demonstrate that the proposed scheme is superior (in terms of accuracy) and more flexible (in adaptation to various systems) than the existing schemes.

\bibliographystyle{IEEEbib}
\bibliography{references}

\begin{thebibliography}{10}

\bibitem{Songze2020}
Songze Li and Salman Avestimehr,
\newblock ``Coded computing: Mitigating fundamental bottlenecks in large-scale
  distributed computing and machine learning,''
\newblock {\em Foundations and Trends in Communications and Information
  Theory}, vol. 17, no. 1, pp. 66--96, 2020.

\bibitem{yu2017polynomial}
Qian Yu, Mohammad Maddah-Ali, and Salman Avestimehr,
\newblock ``Polynomial codes: an optimal design for high-dimensional coded
  matrix multiplication,''
\newblock {\em Advances in Neural Information Processing Systems}, vol. 30,
  2017.

\bibitem{dutta2019optimal}
Sanghamitra Dutta, Mohammad Fahim, Farzin Haddadpour, Haewon Jeong, Viveck
  Cadambe, and Pulkit Grover,
\newblock ``On the optimal recovery threshold of coded matrix multiplication,''
\newblock {\em IEEE Transactions on Information Theory}, vol. 66, no. 1, pp.
  278--301, 2019.

\bibitem{gupta2018oversketch}
Vipul Gupta, Shusen Wang, Thomas Courtade, and Kannan Ramchandran,
\newblock ``Oversketch: Approximate matrix multiplication for the cloud,''
\newblock in {\em 2018 IEEE International Conference on Big Data (Big Data)}.
  IEEE, 2018, pp. 298--304.

\bibitem{jahani2021codedsketch}
Tayyebeh Jahani-Nezhad and Mohammad~Ali Maddah-Ali,
\newblock ``Codedsketch: A coding scheme for distributed computation of
  approximated matrix multiplication,''
\newblock {\em IEEE Transactions on Information Theory}, vol. 67, no. 6, pp.
  4185--4196, 2021.

\bibitem{yu2019lagrange}
Qian Yu, Songze Li, Netanel Raviv, Seyed Mohammadreza~Mousavi Kalan, Mahdi
  Soltanolkotabi, and Salman~A Avestimehr,
\newblock ``Lagrange coded computing: Optimal design for resiliency, security,
  and privacy,''
\newblock in {\em The 22nd International Conference on Artificial Intelligence
  and Statistics}. PMLR, 2019, pp. 1215--1225.

\bibitem{chen2018draco}
Lingjiao Chen, Hongyi Wang, Zachary Charles, and Dimitris Papailiopoulos,
\newblock ``Draco: Byzantine-resilient distributed training via redundant
  gradients,''
\newblock in {\em International Conference on Machine Learning}. PMLR, 2018,
  pp. 903--912.

\bibitem{amiri2019computation}
Mohammad~Mohammadi Amiri and Deniz G{\"u}nd{\"u}z,
\newblock ``Computation scheduling for distributed machine learning with
  straggling workers,''
\newblock {\em IEEE Transactions on Signal Processing}, vol. 67, no. 24, pp.
  6270--6284, 2019.

\bibitem{kosaian2018learning}
Jack Kosaian, KV~Rashmi, and Shivaram Venkataraman,
\newblock ``Learning a code: Machine learning for approximate non-linear coded
  computation,''
\newblock {\em arXiv preprint arXiv:1806.01259}, 2018.

\bibitem{kosaian2020learning}
Jack Kosaian, KV~Rashmi, and Shivaram Venkataraman,
\newblock ``Learning-based coded computation,''
\newblock {\em IEEE Journal on Selected Areas in Information Theory}, vol. 1,
  no. 1, pp. 227--236, 2020.

\bibitem{agrawal2022learning}
Navneet Agrawal, Yuqin Qiu, Matthias Frey, Igor Bjelakovic, Setareh Maghsudi,
  Slawomir Stanczak, and Jingge Zhu,
\newblock ``A learning-based approach to approximate coded computation,''
\newblock in {\em 2022 IEEE Information Theory Workshop (ITW)}. IEEE, 2022, pp.
  600--605.

\bibitem{xiao2017fashion}
Han Xiao, Kashif Rasul, and Roland Vollgraf,
\newblock ``Fashion-mnist: a novel image dataset for benchmarking machine
  learning algorithms,''
\newblock {\em arXiv preprint arXiv:1708.07747}, 2017.

\bibitem{kadam2020cnn}
Shivam~S Kadam, Amol~C Adamuthe, and Ashwini~B Patil,
\newblock ``Cnn model for image classification on mnist and fashion-mnist
  dataset,''
\newblock {\em Journal of scientific research}, vol. 64, no. 2, pp. 374--384,
  2020.

\bibitem{yu2015multi}
Fisher Yu and Vladlen Koltun,
\newblock ``Multi-scale context aggregation by dilated convolutions,''
\newblock {\em arXiv preprint arXiv:1511.07122}, 2015.

\end{thebibliography}
\end{document}